\documentstyle[prl,aps]{revtex}
\begin{document}
\draft

\title{Jellium model of metallic nanocohesion}
\author{C.~A.~Stafford,$^{1,*}$ D.\ Baeriswyl,$^1$ and J.\ B\"urki$^{1,2}$} 
\address{$\mbox{}^1$Institut de Physique 
Th\'{e}orique, Universit\'{e} de Fribourg,
CH-1700 Fribourg, Switzerland}
\address{$\mbox{}^2$Institut Romand de Recherche Num\'erique en 
Physique des Mat\'eriaux
CH-1015 Lausanne}
\date{Received by Phys.Rev.Lett. 17 March 1997}
\maketitle

\begin{abstract}
A unified treatment of the cohesive and conducting properties of metallic
nanostructures 
in terms of the electronic scattering matrix is developed.
A simple picture of metallic nanocohesion in which conductance channels
act as delocalized chemical bonds is derived in the jellium approximation.
Universal force oscillations 
of order $\varepsilon_F/\lambda_F$ are predicted when a metallic quantum wire 
is stretched to the breaking point, which are synchronized
with quantized jumps in the conductance. 

\end{abstract}


Cohesion in metals is due to the formation of bands,
which arise from the overlap of atomic orbitals.
In a metallic constriction with nanoscopic cross section, the transverse
motion is quantized, leading to a finite number of subbands below the
Fermi energy $\varepsilon_F$.  A striking consequence of these discrete
subbands is the phenomenon of conductance quantization \cite{quantization}.
The cohesion in a metallic nanoconstriction must also
be provided by these discrete subbands, which may be thought of as chemical
bonds which are delocalized over the cross section.
In this Letter, we confirm this intuitive picture of metallic nanocohesion 
using a simple jellium model.  Universal force oscillations of order
$\varepsilon_F/\lambda_F$ are predicted in metallic nanostructures exhibiting
conductance quantization, where $\lambda_F$ is the Fermi wavelength.  
Our results are in quantitative agreement with
the recent pioneering experiment of 
Rubio, Agra\"{\i}t, and Vieira \cite{nanoforce}, who
measured simultaneously the force and conductance during
the formation and rupture of an atomic-scale Au contact.
Similar experimental results have been obtained independently by
Stalder and D\"urig \cite{nanoforce2}.

Quantum-size effects on the mechanical properties of metallic systems have
previously been observed in ultrasmall metal clusters \cite{clusters},
which exhibit enhanced stability for certain {\em magic numbers} of atoms.
These magic numbers have been rather well explained in terms of a shell
model based on the jellium approximation \cite{clusters}.  
The success
of the jellium approximation in these closed nanoscopic systems motivates
its application to open (infinite) systems, which are the subject of 
interest here.  We investigate the conducting and mechanical
properties of a nanoscopic constriction connecting two macroscopic
metallic reservoirs.  
The natural framework in which to investigate 
such an open system is the scattering approach developed by
Landauer \cite{landauer} and B\"uttiker \cite{condtheory}.
Here, we extend the formalism of Ref.\ \cite{condtheory}, which 
describes electrical conduction, to describe the mechanical
properties of a confined electron gas as well.

\begin{figure}
\vbox to 15cm {\vss\hbox to 17cm
  {\hss\
    {\includegraphics{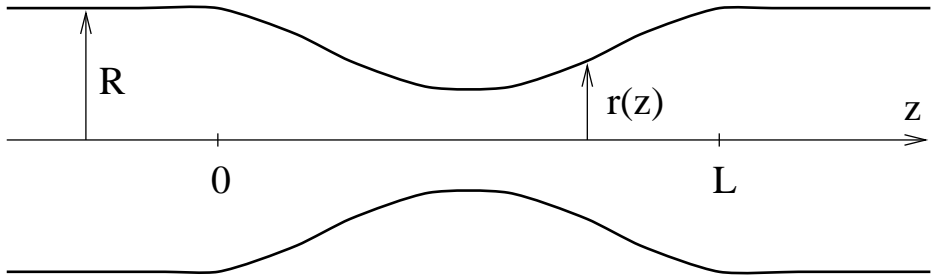}}
   \hss}
}
\caption{
Schematic diagram of a constriction in a cylindrical
quantum wire. Electrons are confined along the $z$ axis by a hard-wall
potential at $r=r(z)$.  Two
different geometries are considered:  $r(z)=(R+R_{\rm min})/2 + (R-
R_{\rm min})\cos(2\pi z/L)/2$ (cosine constriction) and $r(z)=R_{\rm min}
+(R-R_{\rm min})(2z/L-1)^2$ (parabolic constriction), with $r(z)=R$ for
$z<0$ and $z>L$. The minimum radius
of the neck $R_{\rm min}$ as a function of the elongation $\Delta L/L_0$
is determined by a constant volume constraint $\int_0^L r(z)^2dz
=R^2 L_0$.}
\label{fig.geometry}
\end{figure}

For definiteness, we consider a constriction of length
$L$ in an infinitely long cylindrical wire of radius $R$, 
as shown in Fig.\ \ref{fig.geometry}.  
We neglect electron-electron interactions, and assume the electrons to be
confined along the $z$ axis by a hard-wall potential at $r=r(z)$.
This model is considerably simpler than a self-consistent jellium
calculation \cite{clusters}, but should suffice to capture the essential
physics of the problem.
Outside the constriction, the Schr\"odinger equation is separable, and the
scattering states can be written as
\begin{equation}
\psi_{kmn}^{\pm}(\phi,r,z) = 
e^{\pm i kz+im\phi} J_m(\gamma_{mn}r/R),
\label{inoutstates}
\end{equation}
where the quantum numbers $\gamma_{mn}$ are the roots of the Bessel
functions $J_m(\gamma_{mn})=0$.  These scattering states may be
grouped into subbands characterized by the quantum numbers $m$ and 
$n$, and we shall use the notation $\nu=(m,n)$.  The 
energy of an electron in subband $\nu$ is 
$\varepsilon(k)=\varepsilon_{\nu}+ \hbar^2k^2/2m$, where 
\begin{equation}
\varepsilon_{\nu} = \frac{\hbar^2 \gamma_{\nu}^2}{2mR^2}.
\label{e.nu}
\end{equation}
The fundamental theoretical
quantity is the scattering matrix of the constriction
$S(E)$, which connects the incoming and outgoing scattering states.
For a two-terminal device, such as that shown
in Fig.\ \ref{fig.geometry}, $S(E)$ can be decomposed into four submatrices
$S_{\alpha\beta}(E)$, $\alpha$, $\beta=1,2$, where 1 (2) indicates scattering
states to the left (right) of the constriction.  Each submatrix $S_{\alpha
\beta}(E)$ is a matrix in the scattering channels $\nu\nu'$.

In terms of the scattering matrix, 
the electrical conductance is given by \cite{condtheory}
\begin{equation}
G = \frac{2e^2}{h} \int dE\, \frac{-df}{dE} \mbox{Tr}\left\{
S_{12}^{\dagger}(E) S_{12}(E)\right\},
\label{condformula}
\end{equation}
where $f(E)=\{\exp[\beta(E-\mu)]+1\}^{-1}$ is the Fermi distribution function
and a factor of 2 has been included to account for spin degeneracy.
The grand canonical potential of the system is
\begin{equation}
\Omega = -k_B T \int dE\, D(E) 
\ln\left(1+e^{-\beta(E-\mu)}\right),
\label{gibbs1}
\end{equation}
where the density of states in the constriction may be
expressed in terms of the scattering matrix as \cite{Iopen,partial.dos}
\begin{equation}
D(E) = \frac{1}{2\pi i} \sum_{\alpha,\beta} \mbox{Tr} \left\{
S_{\alpha\beta}^{\dagger}(E) \frac{\partial S_{\alpha\beta}}{
\partial E}
- S_{\alpha\beta}(E) 
\frac{\partial S_{\alpha\beta}^{\dagger}}{\partial E}
\right\}.
\label{dos}
\end{equation}
Eqs.\ (\ref{condformula}) to (\ref{dos}) allow one to treat the conducting and
mechanical properties of a confined electron gas on an equal footing,
and provide the starting point for our calculation.

We are interested in the mechanical properties of a metallic nanoconstriction
in the regime of conductance quantization.  The necessary condition to have
well-defined conductance plateaus in a three-dimensional
constriction was shown by Torres, Pascual, and S\'aenz \cite{torres} to be
$(dr/dz)^2 \ll 1$.  In this limit, Eqs.\ (\ref{condformula}) to (\ref{dos})
simplify considerably because one may employ the adiabatic approximation
\cite{goldstein}.  In the adiabatic limit, the transverse motion is separable
from the motion parallel to the $z$ axis, so 
Eqs.\ (\ref{inoutstates}) and (\ref{e.nu}) remain valid in the region of the 
constriction, with $R$ replaced by $r(z)$.  The channel energies
thus become functions of $z$, $\varepsilon_{\nu}(z)=\hbar^2 \gamma_{\nu}^2/
2mr(z)^2$.  In this limit, the scattering matrices
$S_{\alpha\beta}(E)$, $\alpha,\,\beta=1,\,2$ are 
diagonal in the channel indices, 
leading to an effective one-dimensional scattering problem.
The condition $(dr/dz)^2 \ll 1$ and the requirement that
the radius of the wire outside the constriction not be smaller
than an atomic radius (i.e., $k_F R > 1$) 
automatically imply the validity of the WKB approximation.

Since the energy differences between the transverse channels in an atomic-scale 
constriction are large compared to $k_B T$ at ambient temperature, we
restrict consideration in the following to the case $T=0$.
In the adiabatic approximation, the conductance becomes
\begin{equation}
G=\frac{2e^2}{h}\sum_{\nu}T_{\nu},
\label{condform2}
\end{equation}
where the transmission probability for channel $\nu$ 
may be calculated
using a variant of the WKB approximation \cite{glazman,brandbyge}, which 
correctly describes the rounding of the conductance steps at threshold.
The density of states in the constriction in the adiabatic
approximation is
\begin{equation}
D(E)=\frac{2}{\pi} \sum_{\nu} \frac{d\Theta_{\nu}}{dE},
\label{dnde}
\end{equation}
where the total phase shift is given in the WKB approximation by
\begin{equation}
\Theta_{\nu}(E)=\left(2m/\hbar^2\right)^{1/2}
\int_0^L dz \, [E-\varepsilon_{\nu}(z)]^{1/2},
\label{theta}
\end{equation}
the integral being restricted to the region where $\varepsilon_{\nu}(z)< E$.
The grand canonical potential of the system is thus
\begin{equation}
\Omega = -\frac{8\varepsilon_F}{3\lambda_F} 
\int_0^L dz \, \sum_{\nu} \mbox{}^{'}
\left(1-\frac{\varepsilon_{\nu}(z)}{\varepsilon_F}\right)^{3/2},
\label{gibbs2}
\end{equation}
the sum being over channels with $\varepsilon_{\nu}(z) < \varepsilon_F$.
Under elongation, the tensile force is given by $F=-\partial \Omega/
\partial L$.  It is easy to show that $F$ is invariant under a 
stretching of the geometry $r(z) \rightarrow r(\lambda z)$, i.e.,
\begin{equation}
F=\frac{\varepsilon_F}{\lambda_F}  f(\Delta L/L_0, k_F R),
\label{scaling1}
\end{equation}
where $f(x,y)$ is a dimensionless function.  Nonuniversal corrections
to $F$ occur in very short
constrictions, for which the adiabatic approximation breaks down. The
leading order correction to the integrand in Eq.\ (\ref{gibbs2}) is
$-(3\pi/64) k_F r(z) (dr/dz)^2$, leading to a relative error in $F$ of 
$\sim 2\sin^2\theta/4$, where $\theta$ is the opening angle of the 
constriction.  Using a modified Sharvin equation \cite{torres} to estimate
the diameter of the contact versus elongation 
for the experiment of Ref.\ \cite{nanoforce}
indicates an opening angle $\theta \lesssim 45^{\circ}$,
for which the nonuniversal corrections are 
$\lesssim 8\%$, justifying the above approach.

\begin{figure}
\vbox to 19cm {\vss\hbox to 17cm
  {\hss\
    {\includegraphics{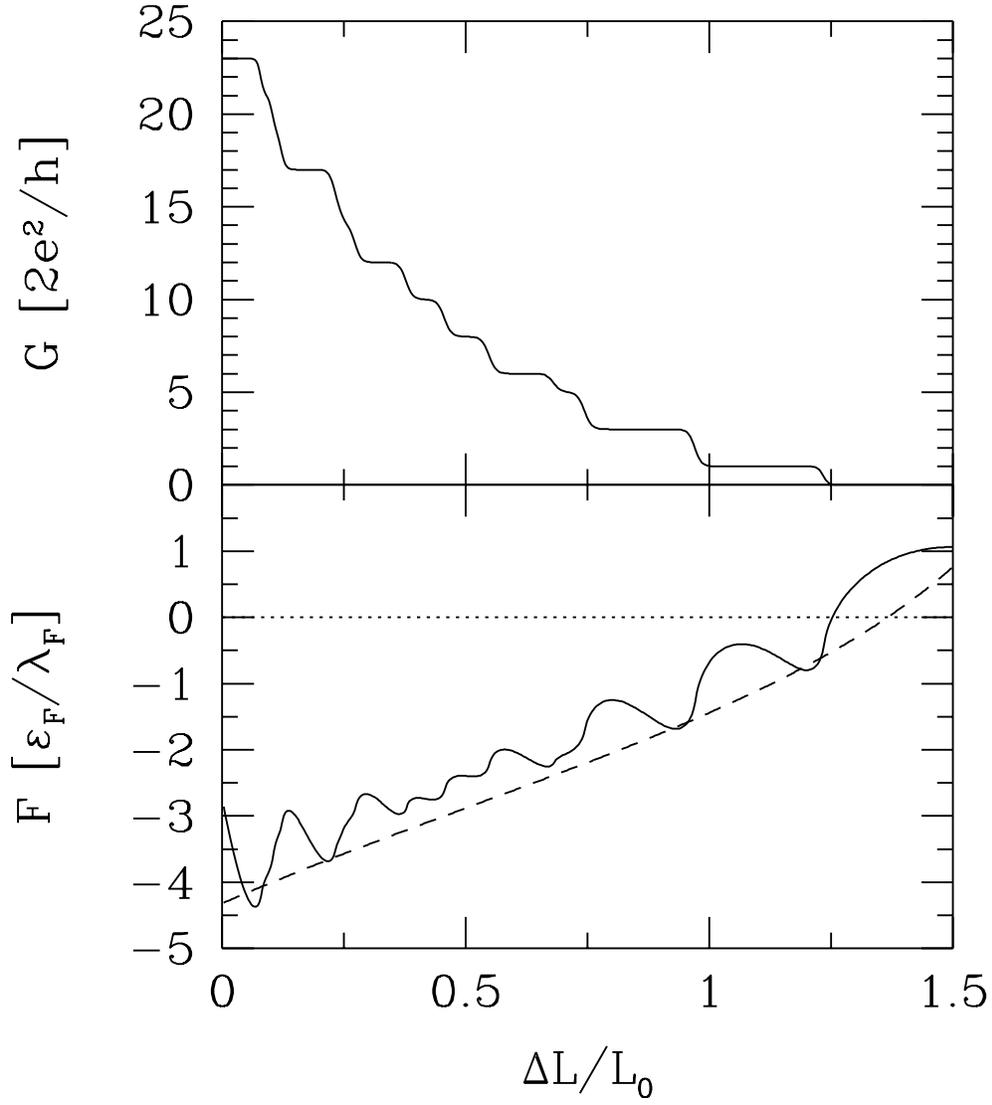}}
   \hss}
}
\caption{
Electrical conductance $G$ and tensile force $F$ of a cosine
constriction in a cylindrical quantum wire of radius $k_F R=11$
versus the elongation $\Delta L/L_0$.  For the calculation of $G$, an
initial length $k_FL_0=50$ was assumed.  The dashed line indicates the 
contribution to the force due to the macroscopic 
surface tension $F_S=-\sigma \partial S/\partial L$,
where $S$ is the surface area of the system and $\sigma=\varepsilon_F 
k_F^2/16\pi$.
$F_S$ determines the overall slope of $F$, on which are superimposed
the quantum oscillations due to the discrete transverse channels.
}
\label{fig.fcond}
\end{figure}

Fig.\ \ref{fig.fcond} shows the conductance and force of a metallic
nanoconstriction as a function of the elongation, calculated from Eqs.\
(\ref{condform2}) and (\ref{gibbs2}).  Here an ideal plastic deformation was
assumed, i.e., the volume of the constriction was held constant \cite{plastic}. 
The correlations between the force and the conductance are striking:  $|F|$
increases along the conductance plateaus, and 
decreases sharply when the conductance drops.  
The constriction becomes unstable when the last conductance channel is cut off.
Some transverse channels 
are quite closely spaced, and in these cases, the
individual conductance plateaus [{\it e.g.}, $G/(2e^2/h)=14$, 15, 19, 21]
and force oscillations are difficult to resolve.
Fig.\ \ref{fig.fcond} is remarkably similar to the experimental results of 
Refs.\ \onlinecite{nanoforce} and \onlinecite{nanoforce2}, 
both qualitatively and quantitatively.
Inserting the value $\varepsilon_F/\lambda_F \simeq 1.7\mbox{nN}$ for Au, we
see that both the overall scale of the force for a given value of the 
conductance and the heights of the last two force oscillations are in
quantitative agreement with the data shown in Fig.\ 1 of Ref.\ 
\onlinecite{nanoforce}.  
We wish to emphasize that the calculation of $F$ presented
in our Fig.\ \ref{fig.fcond} contains no adjustable parameters 
\cite{parameters}.
The increase of $|F|$ along the conductance plateaus and
the rapid decrease at the conductance steps were described in Ref.\
\cite{nanoforce} as ``elastic'' and ``yielding'' stages, 
respectively.  With our intuitive picture of a conductance
channel as a delocalized metallic bond, it is natural to interpret these
elastic and yielding stages as the stretching and breaking of these bonds.

\begin{figure}
\vbox to 19cm {\vss\hbox to 17cm
  {\hss\
    {\includegraphics{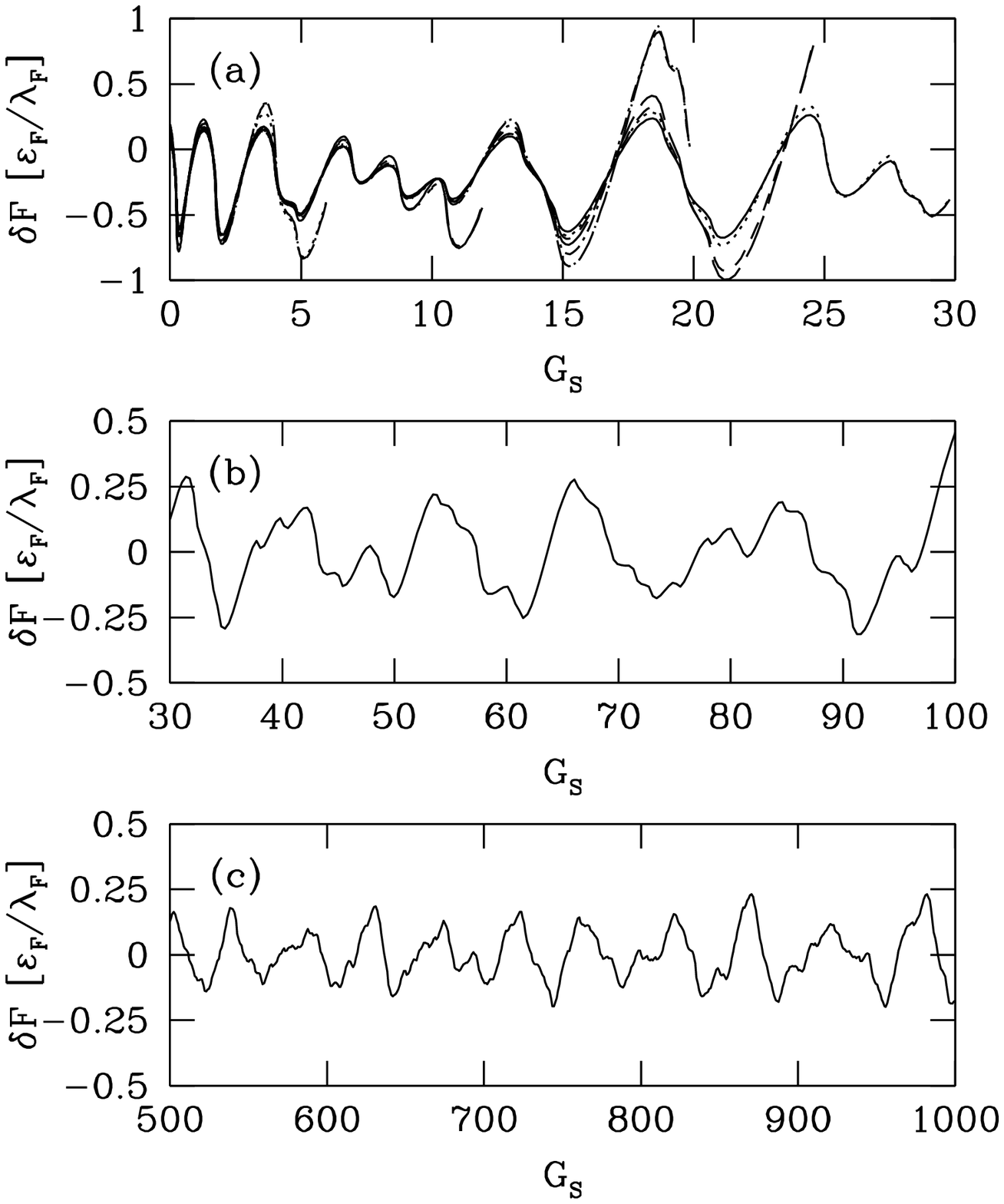}}
   \hss}
}
\caption{Force fluctuations versus the corrected Sharvin conductance.
(a) Results for 
cosine and parabolic constrictions in cylindrical quantum wires with
radii $k_F R=6$, 8, 10, 11, and 12 (left to right).  The results for
the cosine and parabolic constrictions are almost indistinguishable when
plotted as a function of $G_s$,
despite the fact that the total elongation and total work done differ
by roughly a factor of 2 in the two cases.  The dependence on $R$ is
also very weak, except for $R_{\rm min} \approx R$ (the rightmost portion
of each curve).  (b) Force fluctuations
for a cosine constriction in a square quantum wire with 120 conductance 
channels, and (c) for a wire with 1200 conductance channels. 
}
\label{fig.fosc}
\end{figure}

The fluctuations in $F$ due to the discrete transverse channels
may be thought of as arising from finite-size corrections to the surface tension
$\sigma$.  However, as in the case of universal conductance
fluctuations \cite{condfluct}, it is more instructive to consider the 
extensive quantity $F$ itself, rather than the intensive quantity 
$\sigma$.  Approximating the sum in Eq.\ (\ref{gibbs2}) by an integral, and
keeping the leading order corrections, one obtains
\begin{equation}
\Omega=\omega V + \sigma S - \frac{2\varepsilon_F}{3\lambda_F} L 
+ \delta \Omega,
\label{fsize}
\end{equation}
where $V$ is the volume of the system, $S$ is the surface area,
$\omega=-2\varepsilon_F k_F^3/15\pi^2$ is the macroscopic free energy
density, and $\sigma=\varepsilon_F k_F^2/16\pi$ is the macroscopic surface
energy.  The remaining term
$\delta \Omega$ is a quantum correction due to the discrete
transverse channels, and may be either positive or negative.  
Under an ideal plastic deformation, the volume of the
system is unchanged, and the tensile force is
\begin{equation}
F=-\sigma \frac{\partial S}{\partial L} + \frac{2\varepsilon_F}{3\lambda_F}
+ \delta F,
\label{fexp}
\end{equation}
where $\delta F = -\partial (\delta\Omega)/\partial L$.
The first term in Eq.\ (\ref{fexp}) is the contribution to the force 
due to the macroscopic surface tension.  This is plotted as a dashed line
in Fig.\ \ref{fig.fcond}, for comparison.  The macroscopic surface tension
determines the overall slope of $F$.  
The quantum corrections to $F$ due to the
discrete transverse channels consist of a constant term plus the
fluctuating term $\delta F$.  Fig.\ \ref{fig.fosc} shows $\delta F$
for three different geometries and for values of $k_F R$ from 6 to 1200,
plotted versus the corrected Sharvin conductance \cite{torres}
\begin{equation}
G_s = \frac{k_F^2 A_{\rm min} - k_F C_{\rm min}}{4\pi},
\label{sharvin}
\end{equation}
where $A_{\rm min}$ and $C_{\rm min}$ are the area and circumference of the 
constriction at its narrowest point.  $G_s$ gives a smooth 
approximation to $G$.
As shown in Fig.\ \ref{fig.fosc}(a),
the force oscillations obey the approximate scaling relation
\begin{equation}
\delta F(\Delta L/L_0,k_F R) \simeq \frac{\varepsilon_F}{\lambda_F} 
\mbox{Y}(G_s),
\label{scaling2}
\end{equation}
where $Y$ is a dimensionless scaling function which is independent of
the precise geometry $r(z)$.
Eq.\ (\ref{scaling2}) indicates that the force fluctuations, like the
conductance, are dominated by the 
contribution from the narrowest part of the constriction, of radius
$R_{\rm min}$.  The scaling relation (\ref{scaling2}) breaks down when
$R_{\rm min}/R \gtrsim 0.8$.

Fig.\ \ref{fig.fosc} shows that the amplitude of the force fluctuations
persists essentially unchanged to very large values of $G_s$.
It was found to be
\begin{equation}
\Delta Y=\left(\overline{Y^2}-\overline{Y}^2\right)^{1/2} \sim 0.3
\label{dY}
\end{equation}
for $0<G_s\leq 10^4$.  
The detailed functional form of $Y(G_s)$, like
the distribution of widths of the conductance plateaus, depends on the sequence
of quantum numbers $\gamma_{\nu}$, which is determined by the shape of
the cross section.  However, the amplitude of the force fluctuations 
$\Delta Y$ was found to be the same for both circular and square 
cross sections.  Both these geometries are integrable, and hence have
Poissonian distributions of transverse modes.  It is clearly of interest
to investigate the force fluctuations for nonintegrable cross sections, with
non-Poissonian level statistics.

The experiments of Refs.\ \onlinecite{nanoforce,nanoforce2} observed 
well-defined conductance steps, but found no clear evidence of 
conductance quantization for $G/(2e^2/h)>4$ \cite{agrait}.
Deviations of the conductance plateaus from integer values in
metallic point contacts are 
likely to be due to backscattering from imperfections in the lattice
or irregularities in the shape of the constriction \cite{brandbyge}.
We find that such disorder-induced coherent backscattering leads to 
noise-like fine structure \cite{sds.he} in the conductance steps
and force oscillations, with a reduction of the conductance on the
plateaus, but no shift of the overall force oscillations \cite{cas}.
Our prediction of universal force oscillations is consistent with
the experiments of Refs.\ \onlinecite{nanoforce} and
\onlinecite{nanoforce2}, which found force oscillations with
an amplitude comparable to our theoretical prediction for $G/(2e^2/h)$ 
up to 60. 

Molecular dynamics simulations by Landman {\it et al.} 
\cite{landman}, Todorov and Sutton \cite{todorov}, and Brandbyge
{\it et al.} \cite{brandbyge} have suggested that the conductance steps and 
force oscillations
observed in Refs.\ \onlinecite{nanoforce} and \onlinecite{nanoforce2} 
may be due to a sequence of abrupt atomic rearrangements.  
While the discreteness
of the ionic background is not included in the jellium model,
our results nevertheless suggest that 
such atomic rearrangements may be caused by the
breaking of the extended metallic bonds formed by each conductance channel. 
However, it should be emphasized that our prediction of universal
force fluctuations of order $\varepsilon_F/\lambda_F$ 
is not consistent with the simulations of Refs.\ 
\onlinecite{landman} and \onlinecite{todorov}, which predict force fluctuations
which increase with increasing contact area.  This discrepancy may arise
because we consider the equilibrium deformation of a system with extended
electronic wavefunctions, while Refs.\ \onlinecite{brandbyge,landman,todorov}
use a purely 
local interatomic potential and a fast, nonequilibrium
deformation \cite{nonequilibrium}.

In conclusion, we have presented a simple jellium
model of metallic nanocohesion in
which conductance channels act as delocalized metallic bonds.  This model
predicts universal force oscillations of order $\varepsilon_F/\lambda_F$
in metallic nanostructures in 
the regime of conductance quantization, and is able to explain
quantitatively recent experiments on the 
mechanical properties of nanoscopic metallic contacts 
\cite{nanoforce,nanoforce2}.  The formalism developed
here based on the electronic scattering matrix should be applicable 
to a wide variety of problems in the rapidly evolving field of
nanomechanics.

We thank Urs D\"urig for helpful discussions in the early stages of
this work, and for providing us with his results prior to publication.
We have also profited from collaboration with Jean-Luc Barras and Michael
Dzierzawa.  This work was supported in part by Swiss National 
Foundation grant \# 4036-044033.

\begin{references}
\bibitem{plastic} Such a
constraint is natural in the jellium model, but it may be more apt to 
describe elongation than compression in real systems.
\bibitem{parameters} 
The force for a given value of the conductance is essentially
independent of the asymptotic radius
$R$ and of the shape of the constriction [see Fig.\ \ref{fig.fosc}(a)].
\bibitem{nonequilibrium}  Deformation rates of order 1m/s were used in the
simulations of Refs.\
\onlinecite{brandbyge,landman,todorov}, some five orders of magnitude faster 
than in the experiments of Refs.\ \onlinecite{nanoforce,nanoforce2}.
\end{references}
\end{document}